\documentclass[12pt]{iopart}
\usepackage{iopams,setstack,color}
\usepackage{graphicx,braket}
\usepackage[tight]{units}
\newcommand{\RB}{$\mathrm{^{87}Rb}$}
\newcommand{\Xx}{\mathcal{X}}
\newcommand{\Pp}{\mathcal{P}}
\newcommand{\Ww}{\mathcal{W}}
\begin{document}
\title{Optomechanical atom-cavity interaction in the sub-recoil regime}

\author{H Ke\ss ler, J Klinder, M Wolke and A Hemmerich\footnote{Author to whom any correspondence should be addressed.}}
\address{Institut f\"ur Laserphysik, Universit\"at Hamburg, 22761 Hamburg, Germany}
\ead{hemmerich@physnet.uni-hamburg.de}

\begin{abstract}
We study the optomechanical interaction of a Bose-Einstein condensate with a single longitudinal mode of an ultra-high finesse standing wave optical resonator. As a unique feature the resonator combines three extreme regimes, previously not realized together, i.e., strong cooperative coupling, cavity dominated scattering with a Purcell factor far above unity, and sub-recoil resolution provided by a cavity damping rate smaller than four times the single photon recoil frequency. We present experimental observations in good agreement with a two-mode model predicting highly non-linear dynamics with signatures as bistability, hysteresis, persistent oscillations, and superradiant back-scattering instabilities.
\end{abstract}


\maketitle

\section{Introduction}
\label{sec:introduction}

The study of the coupling of single mode radiation to selected electronic degrees of freedom of single atoms has led to the celebrated field of cavity quantum electrodynamics \cite{Rai:01, Wal:06} in the eighties and nineties. In the past decade the focus of research has shifted towards the interaction of a single mode of the radiation field with the external degrees of freedom of well controlled macroscopic objects. Examples of such objects are sub-micron mechanical oscillators like cantilevers or membranes (which has led to the new field of cavity optomechanics \cite{Kip:08, Asp:13}) or, as in this work, droplets of quantum degenerate atomic gases \cite{Rit:13}. When quantum ensembles such as atomic Bose-Einstein condensates (BECs) are considered, a unique arena opens up where the worlds of quantum optics and quantum degenerate many-body physics are brought together in order to prepare and study extreme forms of non-linear quantum matter. The fragile nature of ultracold quantum ensembles 
limits one to dispersive light-matter interactions excluding near resonant excitations followed by spontaneous emission. Such interactions can be engineered in different regimes with respect to three fundamental parameters, which characterize the underlying physics. The regime of cavity dominated scattering arises, if the Purcell factor exceed unity \cite{Pur:46}. In this case scattering into modes of the radiation field not supported by the cavity may be neglected. In the regime of strong cooperative coupling the atom sample acts to shift the cavity transmission resonance by more than its linewidth and hence the back-action of the atoms upon the intra-cavity light field is significant. Finally, the regime of resolved recoil results, if the cavity linewidth is smaller than four times the single photon recoil frequency, which corresponds to the kinetic energy transferred to a resting atom by back-scattering of a single photon. In this case cavity induced back-scattering can only couple a small number of selected momentum states.

In this article, we study the most elementary atom-cavity configuration, providing a maximum of control: a BEC interacting with a single longitudinal mode of a standing wave resonator, which is coupled along the cavity axis by a weak external laser beam. The atom-cavity system operates within the intersection of the three previously mentioned regimes of cavity dominated scattering, strong cooperative coupling, and sub-recoil resolution, and hence despite its elementary character it displays unusual behavior. A striking example is the recent demonstration of sub-recoil cavity cooling \cite{Wol:12}. In this article, we focus on the rich non-linear dynamics arising in the regime of resolved recoil. After a description of our set-up and the experimental protocols, we present experimental observations with signatures as bistability, hysteresis, persistent oscillations, and superradiant back-scattering instabilities. A theoretical model is discussed comprising two particles modes coupled to a single radiation mode, which explains most of our observations.

\section{Experimental set-up}
\label{sec:experimental-setup}
\begin{figure}[htb]
\centering
\includegraphics[scale=1.0, angle=0, origin=c]{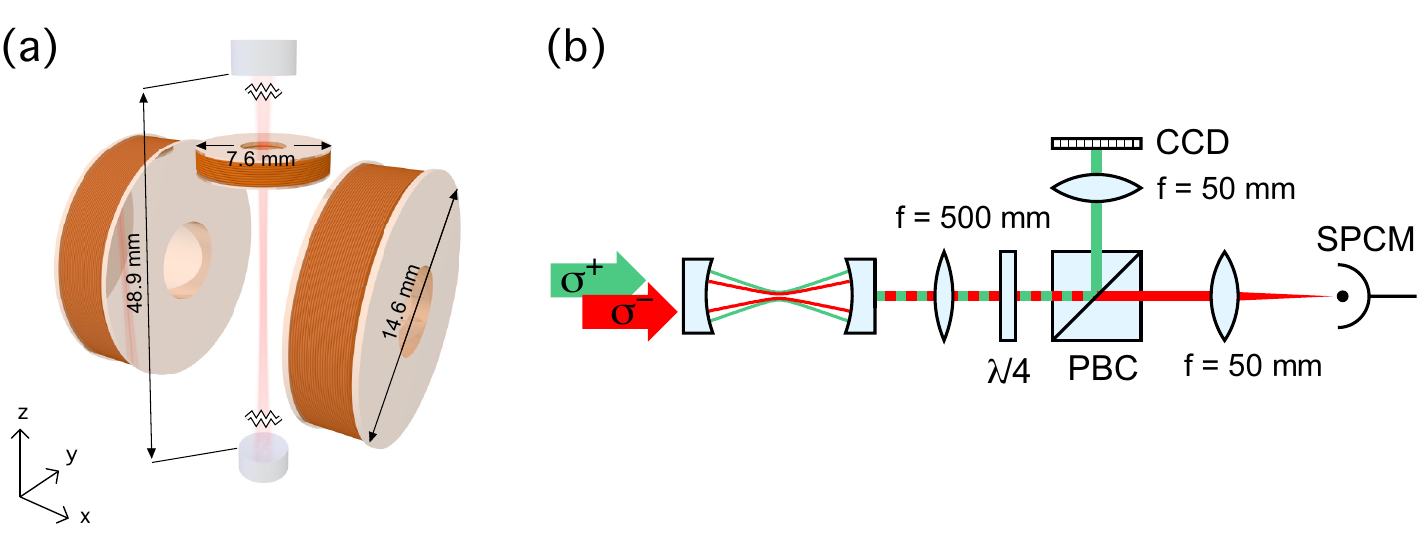}
\caption{Schematic of the experimental set-up. (a) Sketch of the coil system and the mirror set-up. (b) Sketch of cavity coupling scheme with the pump beam (red) and the reference beam (green) shown. After transmission through the cavity the beams are separated by a polarizing beam splitter cube (PBC). The pump beam is detected by a single photon counting module (SPCM), the reference beam is directed to a CCD-camera for monitoring purposes.}
\label{fig:expsetup}
\end{figure}

A cigar-shaped BEC with Thomas-Fermi radii $(3.1, 3.3, 26.8)\,\mu$m and $N = 9\times\,10^4$ - $1.5 \times\,10^5$ \RB-atoms, prepared in the upper hyperfine component of the ground state $\ket{F=2,m_F=2}$, is confined by three centimeter-sized solenoids arranged in a quadrupole Ioffe configuration, thus providing a magnetic trap with a nonzero bias field with trap frequencies $\omega / 2\pi = \unit[(215.6 \times 202.2 \times 25.2)]{Hz}$ \cite{Ess:98, Han:06}. As seen in Fig.~\ref{fig:expsetup}(a), two oppositely mounted mirrors provide a standing wave high finesse cavity integrated into the coil set-up with the cavity axis along the direction of gravity (see table~\ref{tab:cavity} for an overview of the cavity parameters). For the best possible mode match between the atomic ensemble and the cavity mode, the weekly confining $z$-axis of the magnetic trap is aligned with the cavity axis \cite{Kli:10}. All experiments described in the following sections are performed while holding the atoms in this magnetic trap.
\begin{table}[tB]
\caption{\label{tab:cavity} Overview over the cavity parameters for $\lambda=\unit[803]{nm}$.}
\begin{indented}  
\item[]\begin{tabular}{@{}l c c}
\br
Parameter & \multicolumn{2}{c}{Value}\\ 
\mr
Finesse & $\mathcal{F}$ & $ (3.44 \pm 0.05)\times 10^5$ \\
Mode waist &$w_0 $ & $ \unit[(31.2 \pm 0.1)]{\mu m}$ \\
Free spectral range & $\Delta\nu_{\mathrm{FSR}} $ & $ \unit[3\,063]{MHz}$ \\
\begin{minipage}{8em}\setlength{\baselineskip}{0.1ex}Transversal mode frequency distance\end{minipage} & $\Delta\nu_{\mathrm{TEM}}$ & $\unit[(301.5 \pm 0.1)]{MHz}$ \\
Mirror separation & $L $ & $ \unit[48.93]{mm}$\\
Mirror curvature radius & $ R_M $ & $\unit[25.06]{mm} $ \\
Field decay rate & $\kappa $ & $ 2\pi \times \unit[(4.5 \pm 0.1)]{kHz}$\\
Purcell factor & $\eta_\mathrm{c}$ & $44 \pm 1$ \\
\br
\end{tabular}
\end{indented}
\end{table}

For external pumping a laser is coupled through one of the cavity mirrors (lower mirror in Fig.~\ref{fig:expsetup}(a) to a longitudinal TEM$_{00}$-mode. The narrow cavity linewidth ($\Delta\nu_{\mathrm{FWHM}}\approx\unit[9]{kHz}$) places high demands on the frequency control of the pump laser. Our experiments require to tune its frequency with sub-kilohertz resolution across the resonance frequency of the TEM$_{00}$-mode interacting with the BEC. This is accomplished in two steps: First, a reference laser is locked on resonance with a TEM$_{11}$-mode, which provides a cloverleaf-shaped  transverse profile. This mode exhibits a nodal line at the cavity axis such that the interaction with the BEC, which is positioned well in the center of the TEM$_{00}$-mode, is minimized. In the second step, the pump laser, matched to couple the TEM$_{00}$-mode, is locked with an offset frequency $\omega_{\mathrm{p}} - \omega_{\mathrm{ref}} \approx 2\pi \times\unit[2.5]{GHz}$ to the reference laser. This offset is tunable over a range of about $\pm\unit[1]{MHz}$ such that the vicinity of the resonance frequency of the TEM$_{00}$-mode can be accessed. 

The BEC is placed in the immediate vicinity of the cavity axis such that its coupling to the TEM$_{11}$-mode is suppressed with respect to the TEM$_{00}$-mode by a geometrical factor $\vartheta_{g} \approx 9 \times 10^{-5}$. Both lasers have a wavelength of $\lambda=\unit[803]{nm}$ and are therefore detuned by $\unit[8]{nm}$ to the red side of the $^1\mathrm{S}_{1/2} \rightarrow\, ^2\mathrm{P}_{1/2}$ transition of \RB. The offset field in the magnetic trap and hence the quantization axis is directed parallelly to the cavity axis such that the choice of $\sigma^-$-polarization for the pump beam and $\sigma^+$-polarization for the reference beam yields another suppression factor $\vartheta_{p} \approx 0.43$ for the coupling of the TEM$_{11}$-mode. An important prerequisite is the negligible birefringence of our cavity. The choice of orthogonal polarizations also permits to separate both beams (with better than $90\%$ extinction ratio) after transmission through the cavity by polarization optics (see Fig.~\ref{fig:expsetup}(b). During an experimental run we produce the BEC next to the cavity mode with a distance of a few hundred micrometers. With the help of auxiliary coils (not shown in Fig.~\ref{fig:expsetup}) we then transport the atoms midway between the lobes of the TEM$_{11}$-mode into the center of the TEM$_{00}$-mode.

The high finesse of the cavity ($\mathcal{F} = 3.44 \times 10^5$) together with the narrow beam waist ($w_0 \approx 31.2\, \mu\,$m) yield a Purcell factor $\eta_\mathrm{c} \equiv \frac{24\,\mathcal{F}}{\pi\,k^2 w_0^2} \approx 44$ ($k \equiv 2\pi/\lambda$, and $\lambda=\,$ wavelength of pump beam) and hence only one out of 44 photons impinging upon the intra-cavity atom sample is not scattered into the TEM$_{00}$-mode of the cavity. At the same time this yields strong back-action of the atoms onto the cavity resonance frequency. For a uniform atomic sample (merely giving rise to forward scattering) the resonance frequency is shifted by each atom by an amount $\Delta_0 / 2$ with $\Delta_0 = \frac{1}{2} \eta_\mathrm{c} \kappa \,\Gamma \left( \frac{2}{3 \delta_1} + \frac{1}{3 \delta_2}\, \right)$ and $\delta_{1,2}$ denoting the (negative) pump frequency detunings with respect to the relevant atomic D$_{1,2}$ lines. The fractional prefactors in $\Delta_0$ account for the effective line strengths of the D$_{1}$- and D$_{2}$-line components connecting to the $\ket{F=2, m_f=2}$ ground state for $\sigma^-$-polarization of the pump beam and $\Gamma = 2 \pi \times 6.1$~MHz denotes the decay rate of the $5\mathrm{P}$ state of \RB. In our experiment we have $\Delta_0 \approx - 2\pi \times 0.5$~Hz corresponding to $\Delta_0/ \kappa \approx - 1.1 \times 10^{-4}$. Hence, with $N \approx 4 \times 10^{4}$ atoms the regime of strong cooperative coupling ($N\Delta_0>4\kappa$) is entered.

\section{Theoretical model}
\label{sec:theoretical-model}

\begin{figure}[hbt]
\centering
\includegraphics[scale=0.7, angle=0, origin=c]{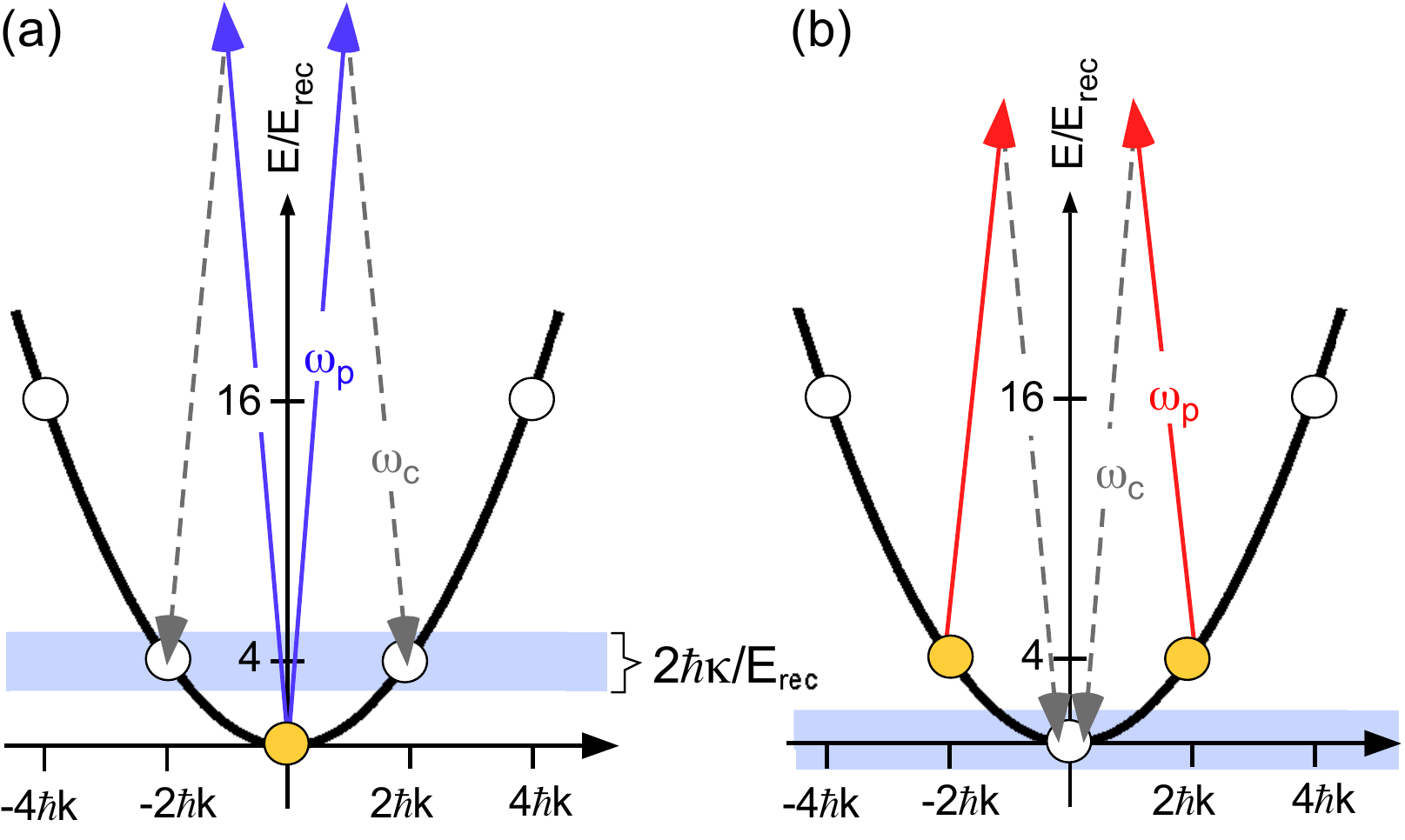}
\caption{Basic back-scattering processes. The momentum states accessible by back-scattering (at multiples of $2\hbar k$) are indicated by disks (orange, if initially populated, white if initially unpopulated). (a) Heating process where the blue detuned ($\omega_{\mathrm{p}} > \omega_{\mathrm{c}}$) pump photons (solid blue arrows) are resonantly back-scattered into the empty cavity (dashed gray arrows). The light blue horizontal bar represents the cavity linewidth. (b) Reversed action, where red detuned ($\omega_{\mathrm{p}} < \omega_{\mathrm{c}}$) pump photons (solid red arrows) are resonantly back-scattered into the cavity mode by atoms initially prepared at $\pm2\hbar k$.}
\label{fig:momentum_transfer}
\end{figure}

The unique feature of the cavity in this work is the combination of a Purcell factor $\eta_\mathrm{c} \approx 44$ well exceeding unity with a cavity field decay rate well below four times the recoil frequency $\kappa < 4\omega_{\mathrm{rec}}$ (where $\omega_{\mathrm{rec}} \equiv \hbar k^2/2 m$ and $m =$ atomic mass. As a consequence, we may completely neglect scattering of pump photons into modes other than the cavity mode and cavity assisted back-scattering can only occur in a narrow resonance window such that only very few motional states are dynamically coupled. The latter is sketched in Fig.~\ref{fig:momentum_transfer} for the simplified case when the possible formation of an intra-cavity optical lattice is neglected and hence the atoms are considered as freely moving rather than populating Bloch states. Consider (see (a)) an atom at zero momentum. Back-scattering of a photon requires that the atom receives $2 \hbar k$ momentum and hence a kinetic energy of $4 \hbar \omega_{\mathrm{rec}}$. Because of the narrow cavity linewidth (indicated by the blue bar in (a)) this process only becomes possible, if the frequency of the pump photon is detuned to the blue side of the cavity resonance by $4\omega_{\mathrm{rec}}$. After the photon emitted into the cavity has left, a second back-scattering process involving the same atom is not supported by the cavity. Hence, back-scattering irreversibly populates the momentum states $\ket{\pm 2 \hbar k}$. Similarly, if initially atoms in the $\ket{\pm 2 \hbar k}$ states are prepared, upon offering photons detuned to the red side of the cavity resonance by $4 \hbar \omega_{\mathrm{rec}}$ these atoms are transferred back to the zero momentum state (see (b)). By applying a suitable sequence of light pulses, with frequencies approaching the cavity resonance from the red side, it should be possible to cool thermal atoms to less than the recoil temperature, in analogy to conventional Raman cooling, albeit at largely increased atomic densities and without the need of spontaneous photons, such that even quantum degeneracy could be reached \cite{San:13}. 

\subsection{Two-mode model}
\label{sec:two-mode-model}
In this section, we outline the theoretical framework used to model our atom-cavity system.  The many-body Hamiltonian reads
\begin{eqnarray}
\label{eq:hamilton-atom}
H_A = \int d^3r \; \hat{\Psi}^{\dagger}(\vec{r},t)  \Big[-\frac{\hbar^2}{2m} \Delta &+& \hat{a}^{\dagger}\hat{a}\; \hbar \Delta_0 \cos^2 (k z) \,+\, V(\vec{r}) 
\\  \nonumber
&+& \,\,\frac{g}{2}\, \hat{\Psi}^{\dagger}(\vec{r},t)\,\hat{\Psi}(\vec{r},t) \Big] \hat{\Psi}(\vec{r},t) \, ,
\end{eqnarray}
where $\Delta_0$ is the light shift per photon, $V(\vec{r})$ with $\vec{r}=(x,y,z)$ denotes the external trap potential,  $\hat{\Psi}(\vec{r},t)$ annihilates a particle at position $\vec{r}$ and time $t$ and $\hat{a}^{\dagger}$ ($\hat{a}$) is the creation (annihilation) operator for a photon in the cavity mode with the spatial field distribution given by $\cos(kz)$ (fixed by the mirror positions). The interaction parameter is $g \equiv 4\pi \hbar^2 a_s / m$ with the $s$-wave scattering length $a_s$ (= 5.7 nm) and the atomic mass $m$. Since the atomic sample is prepared in a strongly elongated external trap aligned with the cavity axis ($z$-axis, cf. Sec. \ref{sec:experimental-setup}), we neglect the weak confinement of the external trap in the $z$-direction, writing $V(\vec{r}) = V(x,y)$. Because the extreme energy selectivity of the cavity acts to exclusively couple the BEC in its zero momentum state to the two excited states with momenta $\pm 2 \hbar k$, the atomic dynamics may be described in terms of two modes $\ket{0}$ and $\ket{\cos(2kz)}$ with spatial wave functions proportional to $\phi(x,y)$ and $\phi(x,y) \cos(2kz)$, respectively, with $|\phi(x,y)|^2$ denoting the normalized ($1 = \int{dxdy |\phi(x,y)|^2}$) transversal density distribution, fixed by the external potential $V(x,y)$. The many-body field operator then takes the simplified form 
\begin{equation}
\label{eq:2modemodel}
\hat{\Psi}(z,t) =  \phi(x,y) \left( \hat{b}_0(t) + \sqrt{2}\, \hat{b}_2(t) \cos(2kz)\right) \mathcal{L}^{-1/2}  \, , 
\end{equation}
where $\mathcal{L} = \int{dz}$ denotes the one-dimensional mode volume, and $\hat{b}_{\nu}$ ($\nu \in \{0,2\}$), which describe the annihilation of particles in the motional states $\ket{0}$ and $\ket{\cos(2kz)}$, satisfy the bosonic commutation relations $[\hat{b}_{\nu}(t), \hat{b}^{\dagger}_{\mu}(t)] = \delta_{\nu \mu}$. Similar dual-mode expansions have been previously employed in studies with cavities exhibiting linewidths exceeding the recoil frequency by several hundreds, where it represents an approximation only valid for very small admixtures of the $\cos(2kz)$-mode \cite{Bre:08}. In contrast, in this work, our analysis in terms of two particle modes is justified even for complete inversion, i.e., if $\hat{b}_0(t) = 0$. The intra-cavity light field is described by the Hamiltonian
\begin{equation}
\label{eq:hamilton-light}
\frac{1}{\hbar} H_{L} = -\delta_c \, \hat{a}^{\dagger}\hat{a}  + i \eta \left( \hat{a}^{\dagger} - \hat{a}   \right)
\end{equation}
with the cavity-pump detuning $\delta_c \equiv \omega_{\mathrm{p}} - \omega_{\mathrm{c}}$ and the cavity pump rate $\eta$. Upon inserting Eq.~(\ref{eq:2modemodel})  into Eq.~(\ref{eq:hamilton-atom}) the total Hamiltonian $H = H_{A}+H_{L}$ is obtained as
\begin{eqnarray}
\label{eq:total-hamiltonian}
\frac{1}{\hbar} H = 4 \omega_{\mathrm{rec}}\, \hat{b}_2^{\dagger}\hat{b}_2  
- \left( \delta_{\mathrm{eff}} - \frac{1}{\sqrt{8}} \Delta_0 \left[\hat{b}_0^{\dagger}\hat{b}_2 + \hat{b}_2^{\dagger}\hat{b}_0 \right] \right) \hat{a}^{\dagger}\hat{a} + i \eta \left( \hat{a}^{\dagger} - \hat{a}   \right) 
\\ \nonumber
+\,\frac{1}{2}\,\tilde{g}_{\mathrm{1D}}\, \left(  \hat{b}_0^{\dagger}\hat{b}_0^{\dagger}\left[\hat{b}_0\hat{b}_0+\hat{b}_2\hat{b}_2\right] 
+ \hat{b}_2^{\dagger}\hat{b}_2^{\dagger}\left[\hat{b}_0\hat{b}_0+\frac{3}{2}\,\hat{b}_2\hat{b}_2\right]
+ 4\hat{b}_2^{\dagger}\hat{b}_0^{\dagger}\hat{b}_0\hat{b}_2    \right) \, ,
\end{eqnarray}
where $\delta_{\mathrm{eff}} \equiv \delta_C - \frac{1}{2} N \Delta_0$ denotes the effective pump detuning with respect to the cavity resonance shifted by the $N$ atoms due to forward scattering, $\tilde{g}_{\mathrm{1D}} = g_{\mathrm{1D}}  \,\mathcal{L}^{-1}$, and $g_{\mathrm{1D}} = g \int{dxdy |\phi(x,y)|^4}$. This Hamiltonian has been previously studied in the limit $\hat{b}_0 \approx \sqrt{N}$ and $\hat{b}_2^{\dagger}\hat{b}_2 \ll N$ such that it reduces to the simpler optomechanical form
\begin{equation}
\label{eq:reduced-hamiltonian}
\frac{1}{\hbar} H =  - \delta_{\mathrm{eff}}\, \hat{a}^{\dagger} \hat{a}
+ \sqrt{\frac{N}{8}}\, \Delta_0 \, \hat{a}^{\dagger} \hat{a} \,\mathcal{Z}
+ i \eta \left( \hat{a}^{\dagger} - \hat{a}   \right)
+ \,\frac{1}{2}\,\tilde{g}_{\mathrm{1D}} \,\hat{b}_0^{\dagger}\hat{b}_0^{\dagger}\hat{b}_0\hat{b}_0
\end{equation}
with the real operator $\mathcal{Z} \equiv \hat{b}_2^{\dagger} + \hat{b}_2$ denoting the amplitude of a harmonic mechanical oscillator.
In a BEC set-up this oscillator describes a collective vibration of the BEC \cite{Mur:08, Bre:08}. It has been shown in Ref.~\cite{Bre:08}
that the Hamiltonian of Eq.~(\ref{eq:reduced-hamiltonian}) can give rise to persistent oscillations despite the unavoidable damping via cavity loss. The same Hamiltonian (with $\tilde{g}_{\mathrm{1D}} = 0$) also arises in other experimental implementations, for example, in the classical optomechanical scenario of a two-mirror cavity with one mirror mounted on a spring \cite{Dor:83, Mey:85, Man:98}.

In the present work we consider the full Hamiltonian of Eq.~(\ref{eq:total-hamiltonian}). The system dynamics is determined by the Heisenberg equations with respect to the operators $\hat{b}_{\nu}, \nu \in \{0,2\}$ and $\hat{a}$. Limiting the discussion to mean field effects we replace $\hat{b}_{\nu}$ and $\hat{a}$ by complex functions $\beta_{\nu}(t)$ and $\alpha(t)$ and define the real quantities $\Xx=\frac{2}{N}\, \mathrm{Re}\left(\beta_0(t)^{\ast} \beta_2(t)\right)$, $\Pp~=~\frac{2}{N}\, \mathrm{Im}\left(\beta_0(t)^{\ast} \beta_2(t)\right)$ and $\Ww = \frac{1}{N}\left( \beta_2(t)^{\ast} \beta_2(t) - \beta_0(t)^{\ast} \beta_0(t) \right)$ to obtain the set of complex non-linear equations
\begin{eqnarray}
\label{eq:Heisenberg}
\fl \qquad
i\,\frac{\partial}{\partial t} \left(\begin{array}{c} \beta_0 \\ \beta_2 \end{array} \right)
= 
\left(\begin{array}{cc}
\frac{1}{2}  \alpha^{\dagger} \alpha \, \Delta_0 + c_0 & \frac{1}{\sqrt{8}}  \alpha^{\dagger}  \alpha \, \Delta_0 + c_0 \Xx \\  
\frac{1}{\sqrt{8}}  \alpha^{\dagger}  \alpha \, \Delta_0 + c_0 \Xx & \frac{1}{2} \alpha^{\dagger} \alpha \, \Delta_0 + 4\omega_{\mathrm{rec}} + \frac{1}{4} c_0 (1+\Ww)
\end{array} \right) 
\left(\begin{array}{c} \beta_0 \\ \beta_2 \end{array} \right) \,,
\end{eqnarray}
and hence the real non-linear Bloch-equation
\begin{equation}
\label{eq:matter}
\frac{\partial}{\partial t} \left(\begin{array}{c} \Xx \\ \Pp \\ \Ww \end{array} \right)
=
- \left(\begin{array}{c} \frac{1}{\sqrt{2}} \alpha^{\dagger} \alpha \, \Delta_0 + 2 c_0 \Xx \\ 0 \\ 4\omega_{\mathrm{rec}} + \frac{1}{4} c_0 (1+\Ww)\end{array} \right)
\times 
\left(\begin{array}{c} \Xx \\ \Pp \\ \Ww \end{array} \right) \, .
\end{equation}
The collision parameter is defined as $c_0 \equiv g_{\mathrm{1D}} \,\rho_{\mathrm{1D}}/ \hbar \omega_{\mathrm{rec}}$ with $\rho_{\mathrm{1D}}$ denoting the line density of the BEC with respect to the $z$-direction. The Bloch vector $\left(\Xx , \Pp , \Ww \right)$ has unity length with $\Xx + i \,\Pp$ denoting the coherence and $\Ww$ denoting the inversion of the two-level system  \{$\ket{0}$, $\ket{\cos(2kz)}$\}. In contrast to conventional Bloch equation dynamics, the field vector $-\left(\frac{1}{\sqrt{2}} \alpha^{\dagger} \alpha \, \Delta_0 + 2 c_0 \Xx , 0 , 4\omega_{\mathrm{rec}} + \frac{1}{4}c_0 (1+\Ww) \right)$ is itself a dynamical quantity. The intra-cavity light field evolves according to
\begin{equation}
\label{eq:light}
i \frac{\partial}{\partial t} \alpha = - \left( \delta_{\mathrm{eff}} - \frac{1}{\sqrt{8}} \, N \Delta_0 \, \Xx  + i \kappa  \right) \alpha +   i \eta \, .
\end{equation}
In addition to the Hamiltonian evolution, a damping term scaling with the field decay rate $\kappa$ is included, which describes cavity losses. Note that Eqs.~(\ref{eq:matter}), (\ref{eq:light}) are also obtained by writing Maxwell's equations for the light field in the slowly varying amplitude approximation and a Gross-Pitaevskii equation for the atoms. 

\subsection{Stationary states}
\label{sec:stationary-states}

\begin{figure}[bt]
\centering
\includegraphics[scale=1.0, angle=0, origin=c]{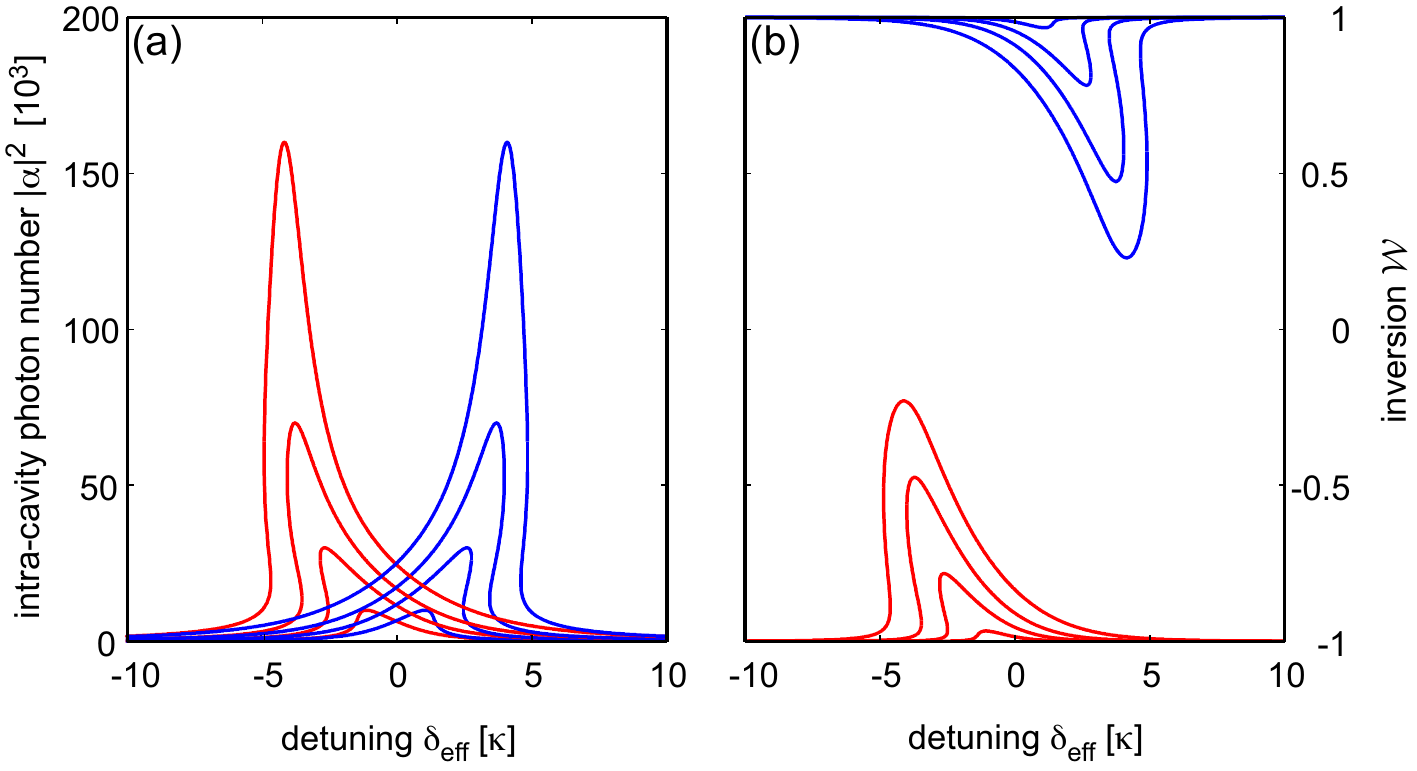}
\caption{Stationary solutions for $N_p = (10, 30, 70, 160) \times 10^3$ and $q=-3.1$. The average intra-cavity photon number $|\alpha|^2$ and the inversion $\Ww$ are plotted versus the effective pump-cavity detuning in (a) and (b), respectively. Blue graphs correspond to $\mathrm{sgn}(\Ww)=+1$ and red curves to $\mathrm{sgn}(\Ww)=-1$.}
\label{fig:steadystate}
\end{figure}
The model defined by Eqs.~(\ref{eq:matter}) and (\ref{eq:light}) exhibits rich non-linear behavior. It provides two types of stationary solutions characterized by the sign of the inversion $\Ww$, which are stable in certain parameter regimes, while in other parameter windows persistent oscillations arise approaching a limit cycle. It is straight forward to determine the stationary solutions by setting the left sides in Eqs.~(\ref{eq:matter}) and (\ref{eq:light}) to zero. We limit the discussion to the interaction free case ($g=0$), however, the gross structure of the results extends to the case of moderate interactions. For better readability we introduce the following abbreviations: the scaled average intra-cavity photon number $\xi \equiv \frac{\Delta_0}{4\omega_{\mathrm{rec}}} \, |\alpha|^2$, the pumping strength parameter $p \equiv \frac{N_p \Delta_0}{4\omega_{\mathrm{rec}}}$ with $N_p \equiv |\eta / \kappa|^2$ denoting the resonant intra-cavity photon number in absence of atoms, and the scaled cooperativity parameter $q \equiv \frac{N \Delta_0}{4 \kappa}$. With these definitions we obtain the relations
\begin{eqnarray}
\label{eq:stationary_states}
p = \xi \left( \left( \frac{\delta_{\mathrm{eff}}}{\kappa} - \mathrm{sgn}(\Ww) \, q\, \frac{\xi}{\left( 1+ \frac{1}{2}\xi^2 \right)^{1/2}}  \right)^2 +1  \right)\,,\;\Ww^2=\frac{1}{1+\frac{1}{2}\xi^2} \, .
\label{eq:stationary_states_b}
\end{eqnarray}
Note that these equations are invariant under a common sign change of $\delta_{\mathrm{eff}}$ and $\Ww$, showing that solutions arise in pairs. Fig.~\ref{fig:steadystate} shows the solutions obtained for different cavity pump strengths $- p = 0.35, 1.04, 2.43, 5.55$ corresponding to $N_p = (10, 30, 70, 160) \times 10^3$ photons in the cavity at the resonance peaks of the intra-cavity photon number plotted in (a). The cooperativity parameter is $q=-3$, which corresponds to $N = 1.1 \times 10^{5}$ atoms in the cavity. In (b) the inversion $\Ww$ corresponding to the graphs in (a) is shown. The sign of $\Ww$ is negative for the red graphs and positive for the blue graphs for arbitrary pump strength. Both types of stationary solutions develop multivalued regions as the pump strength increases, which gives rise to optomechanical bistability and hysteresis. The $\Ww <0$ solutions (red graphs) show a resonance shift towards negative detuning. Recall that dipole forces point towards the intensity maxima for light detuned to the red side of the atomic resonance. Therefore, for $\Ww <0$ solutions increasing intra-cavity intensity yields increased localization of the atoms in the anti-nodes of the intra-cavity optical standing wave and hence increased atom-cavity coupling. Consequently, the intra-cavity optical path length increases and the cavity resonance is shifted towards lower frequencies. The negative shift grows with increasing pump strength and approaches a maximal value $\delta_{\mathrm{eff}} = \frac{1}{\sqrt{8}} N\,\Delta_0$, while the peak value of  $\Ww$ approaches zero, which corresponds to the maximal possible localization. In a band picture the $\Ww <0$ solutions correspond to the stationary scenario, where the atoms are condensed in the lowest band at zero quasi-momentum. The $\Ww >0$ solutions (blue graphs) represent the case of atoms condensed at zero quasi-momentum in the energy minimum of the third band. In this latter case the atomic density scales with $\cos^2(2kx)$ in absence of intra-cavity photons such that the atoms are equally bunched at the minima and the maxima of the cavity mode. In presence of the $-\cos^2(kx)$ lattice potential the atoms are described by the zero quasi-momentum Bloch-function of the third band, which displays reduced density maxima in the antinodes. Hence, for the blue curves, an increased pump strength acts to reduce the coupling with the consequence of a positive line shift. Note also, that the stationary solutions previously studied in the case of weak coupling ($|p| \ll 1$, $\Ww \approx -1$) in Ref.~\cite{Bre:08} correspond to the red solutions in Fig.~\ref{fig:steadystate}.

\subsection{Stability analysis}
\label{sec:stab-cons}
\begin{figure}[btp]
\centering
\includegraphics[scale=0.7, angle=0, origin=c]{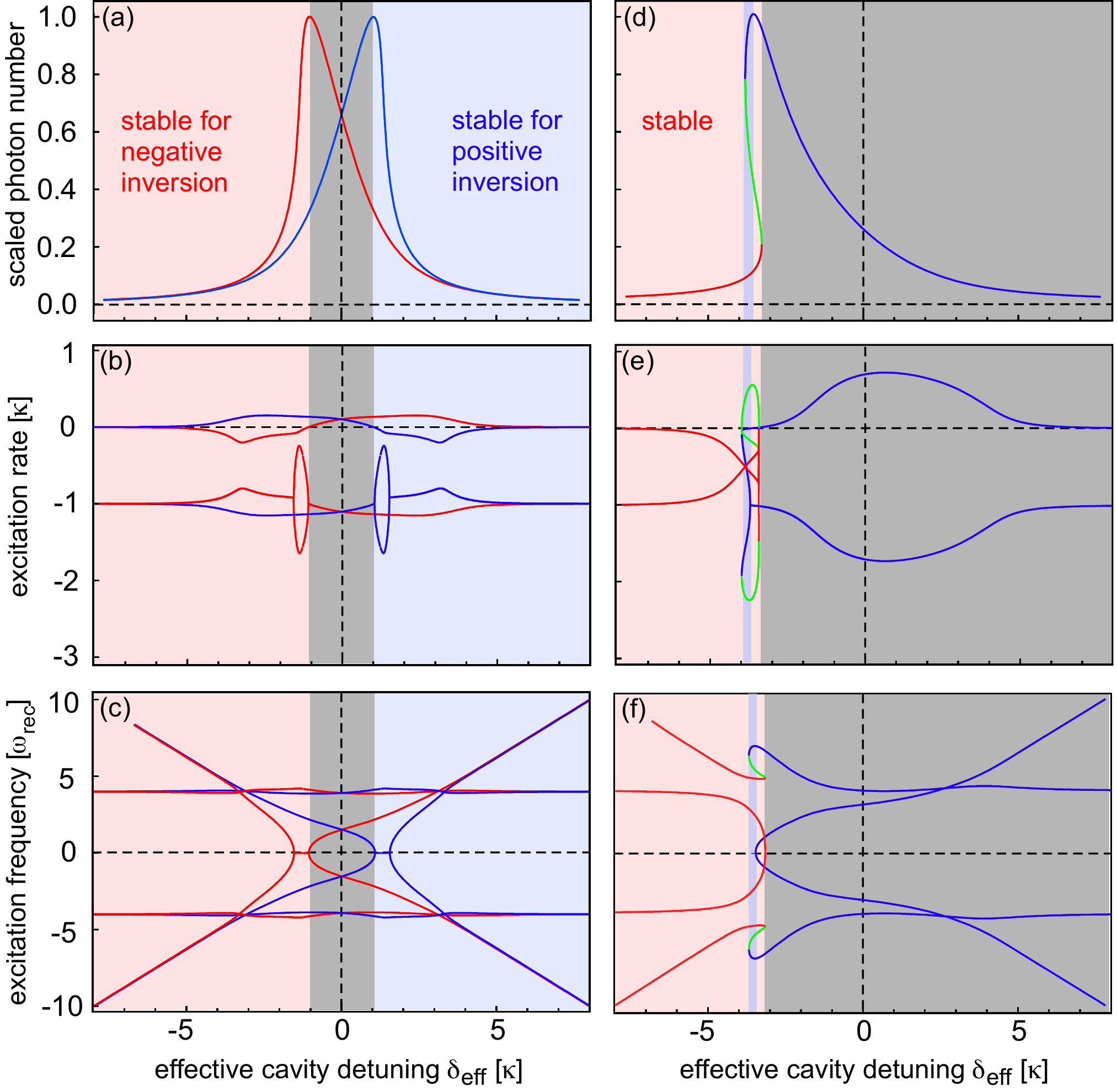}
\caption{Stability analysis for stationary states. The cooperativity parameter is $q=-3$. Left column: $p=-0.4$, right column: $p=-2$. 
 In (a), (b), (c) the (scaled) intra-cavity photon number, the damping rates and the excitation frequencies of both possible stationary solutions ($\Ww <0 \rightarrow$ red traces, $\Ww >0 \rightarrow$ blue traces) of Eq.~(\ref{eq:stationary_states}) are plotted versus $ \delta_{\mathrm{eff}}$. The red and blue areas denote the stability range of the correspondingly colored traces in (a). In the grey areas no stationary solution exists. On the right side (d), (e), (f) are analogue to (a), (b),(c) but for $p=-2$. Here, only the $\Ww <0$ solution is considered for better readability. The colors identify the different branches of the stationary solution in regions, where it becomes multivalued. See text for details.}
\label{fig:stability}
\end{figure}

We now wish to identify those regions, were the stationary solutions of Fig.~\ref{fig:steadystate} are stable. The dynamical equations Eqs.~(\ref{eq:matter}) and (\ref{eq:light}) may be rewritten as $\frac{d}{dt}\vec{\mathcal{Y}} = \vec{\mathcal{F}}(\vec{\mathcal{Y}})$ in terms of the vector $\vec{\mathcal{Y}} =(\mathcal{X},\mathcal{P},\mathcal{W},\mathcal{A},\mathcal{B})$ where $\mathcal{A}$ and $\mathcal{B}$ denote the real and imaginary parts of the field amplitude $\alpha$. The Jacobian $\mathcal{M}(\vec{\mathcal{Y}}_0) \equiv (\partial \mathcal{F}_{\nu}/ \partial \mathcal{Y}_{\mu})_{\vec{\mathcal{Y}}_0}$ evaluated for some stationary solution $\vec{\mathcal{Y}}_0$ determines the stability of $\vec{\mathcal{Y}}_0$. Since $\mathcal{M}$ is real valued, stability of $\vec{\mathcal{Y}}_0$ requires that all eigenvalues of $\mathcal{M}(\vec{\mathcal{Y}}_0)$ have negative real parts, which quantify the damping of excitations along the corresponding eigenvectors. Their imaginary parts yield the excitation frequencies. In Fig.~\ref{fig:stability} the regions of stability are analyzed for $q=-3$ with respect to the effective detuning. In ((a),(b),(c)) weak pumping ($p=-0.4$) is considered. In (a) the intra-cavity intensity is shown for the two possible stationary solutions with negative (red trace) and positive (blue trace) inversion. These curves correspond to the lowest red and blue traces in Fig.~\ref{fig:steadystate} (a), which do not exhibit multivalued regions. Below, the corresponding real (b) and imaginary parts (c) of the eigenvalues of $\mathcal{M}(\vec{\mathcal{Y}}_0)$ are plotted. The regions where the blue or red solutions in (a) become stable are marked by a light blue or light red background. These regions are separated by an area marked in light grey extending between the two resonance maxima in (a). In this range of $\delta_{\mathrm{eff}}$ no stable stationary solution exists. The system rather approaches a periodic trajectory in phase space, i.e., a limit cycle, which is further discussed below. The plots in (d), (e) and (f) correspond to those in (a), (b) and (c), however, for a larger pump strength $p=-2$, such that this stationary solution exhibits  a multivalued region. For simplicity only the solution with negative inversion is plotted. The colors identify the three different branches of the intra-cavity transmission in (d) and the corresponding damping rates and excitation frequencies in (e) and (f), respectively. The region marked by a light red background color indicates stability of that branch of the solution marked in red. The thin blue stripe within the red region indicates that here also the corresponding part of the blue branch is stable. The grey region on the right indicates instability of this solution. Far from the resonance condition $\delta_{\mathrm{eff}} = 0$ the excitation spectra in (c) and (f) show two excited modes for each stationary solution: one with constant energy $4 \hbar \omega_{\mathrm{rec}}$ and one with energy $\hbar \delta_{\mathrm{eff}}$. The former is purely matter-based and corresponds to the excitation of a collective BEC oscillation at $4 \omega_{\mathrm{rec}}$, the latter corresponds to an excitation of the intra-cavity light mode at frequency $\delta_{\mathrm{eff}}$. As is seen from (b) and (e) the photon-like modes are damped at the field decay rate $\kappa$. In the vicinity of $\delta_{\mathrm{eff}} = 0$ matter and photon modes are coupled giving rise to modified excitation frequencies and damping constants.

\begin{figure}[btp]
\centering
\includegraphics[scale=0.7, angle=0, origin=c]{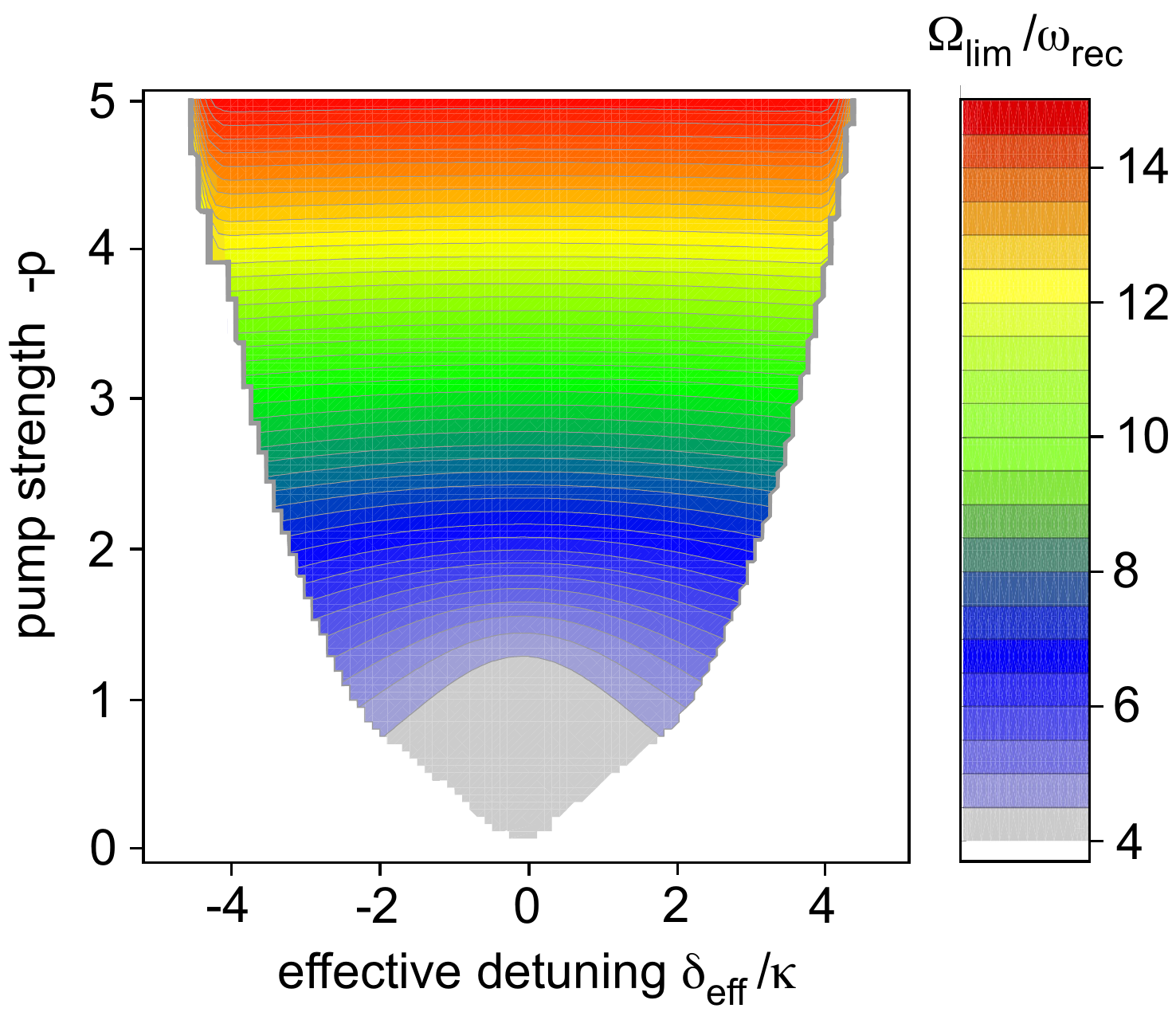}
\caption{The fundamental frequency $\Omega_{\mathrm{lim}}$ of $\Xx(t)$ after a settling time of 100 ms is plotted  versus $\delta_{\mathrm{eff}}$ and $p$ for $q=-3$ and $g=0$.}
\label{fig:limit_cycle}
\end{figure}

Finally, we may also consider the grey area in Fig.~\ref{fig:stability}(a), where the systems approaches a limit cycle with a frequency $\Omega_{\mathrm{lim}}$ determined by $\delta_{\mathrm{eff}}$ and the pump strength $p$. In Fig.~\ref{fig:limit_cycle} we have plotted $\Omega_{\mathrm{lim}}$ versus $\delta_{\mathrm{eff}}$ and $p$ for $q=-3$. The frequency is determined by Fourier-analyzing numerical solutions to Eqs.~(\ref{eq:matter}),~(\ref{eq:light}) after a sufficiently long settling time of 100 ms. It turns out that the same limit cycle is approached independent of the initial conditions. Furthermore, higher harmonic contributions (with frequencies $n\,\Omega_{\mathrm{lim}}, n= 2,3...$) are found to be less than a few percent. Outside the colored region in Fig.~\ref{fig:limit_cycle} one of the stationary solution shown in Fig.~\ref{fig:steadystate} is realized.

\section{Observations and comparison with theory}
\label{sec:observations}

\subsection{Transmission spectroscopy}
\label{sec:transm_spec}

\begin{figure}[bth]
\centering
\includegraphics[scale=0.7, angle=0, origin=c]{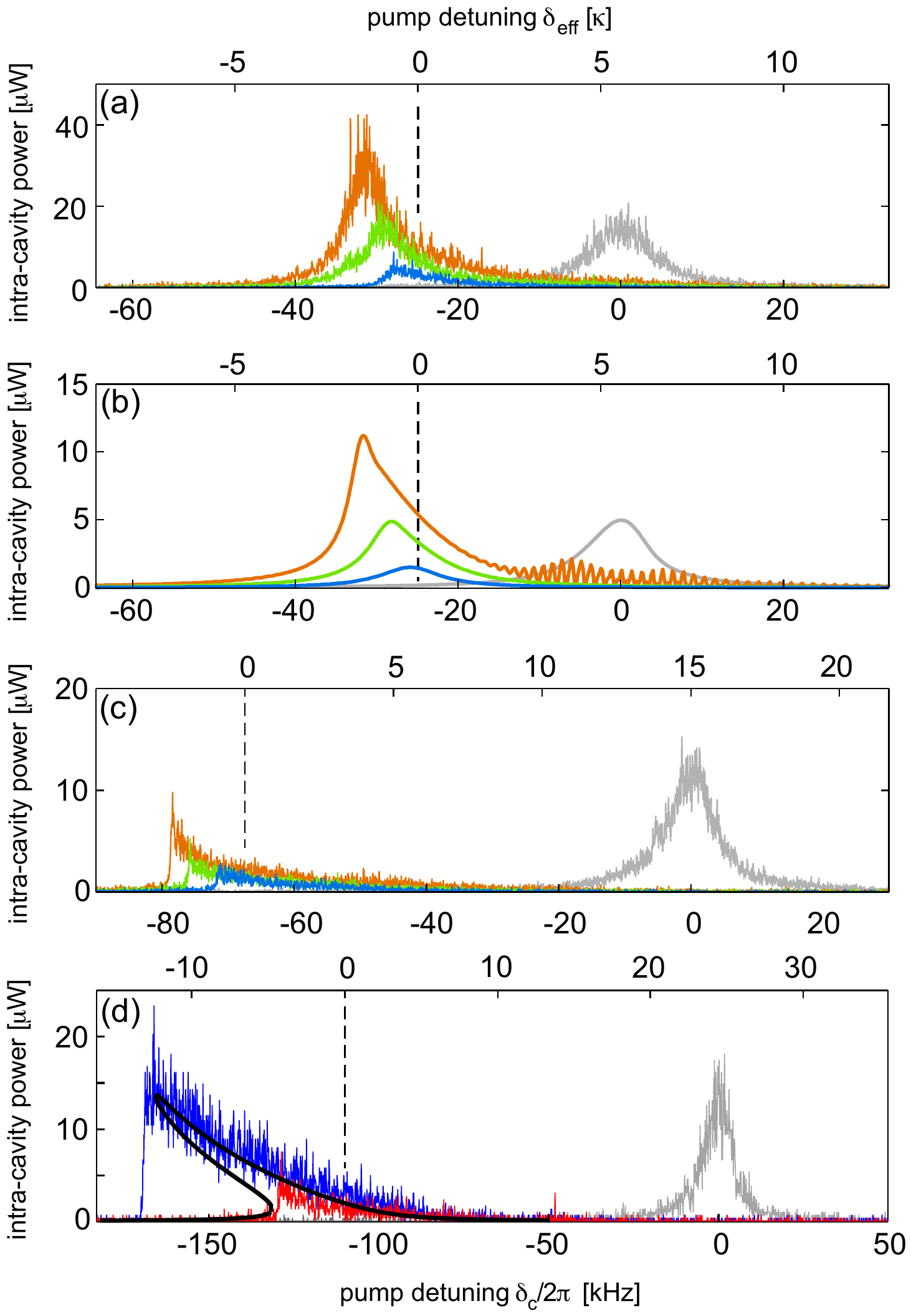}
\caption{Transmission spectroscopy. (a) Transmission spectra for $q = -2.8, N = 10^{5}$ and $p = -0.4, -1.3, -3.2$ (blue, green, orange trace). The positive slope of the frequency scan is $\dot \delta_{\mathrm{eff}}/\kappa^2 \approx 0.07$. (b) Calculations corresponding to (a) using equations Eqs.~(\ref{eq:matter}) and (\ref{eq:light}) with $p = -0.13, -0.43, -1.06$. (c) Spectra for thermal atoms with $q = -8.3, N = 3 \times 10^{5}$ and $p = -0.38, -0.49, -1.04$ (blue, green, orange trace). (d) Spectra for thermal atoms with $q = -12.2$, $N = 4.4 \times 10^{5}$ atoms and $p = -1.47$ for positive (red trace) and negative (blue trace) scanning directions. The black solid trace shows the $\Ww <0$ stationary solution resulting from Eq.~(\ref{eq:stationary_states_b}). The grey trace in each panel shows the spectrum of the empty cavity.}
\label{fig:trans_spec_exp}
\end{figure}

To experimentally probe the steady state solutions of Fig.~\ref{fig:steadystate} (a) we perform transmission spectroscopy of the coupled atom-cavity system. The effective frequency detuning $\delta_{\mathrm{eff}}$ of the pump beam is tuned during $\tau_{\mathrm{scan}}=\unit[10]{ms}$ across a window of $20\,\kappa$ width, which includes the cavity resonance, and the light transmitted through the cavity is recorded. In order to probe stationary solutions the scan duration $\tau_{\mathrm{scan}}$ should be chosen sufficiently long to exceed all dynamical times scales of the coupled atom-cavity system, i.e., $|\dot \delta_{\mathrm{eff}}|/\kappa^2 \ll 1$. Frequency scans with positive slope $\dot \delta_{\mathrm{eff}} > 0$ and with the BEC prepared in the ground state $\ket{0}$ probe the $\Ww<0$ solution (red trace in Fig.~\ref{fig:stability} (a)), starting in its stable region. After the resonance maximum is passed, this solution becomes unstable, however, since the excitation (negative damping) rate (upper red trace in Fig.~\ref{fig:stability} (b)) remains small (below $0.2\, \kappa$) significant deviations from the stationary solution cannot develop, if the scan time is not too long. Adjustment of $\dot \delta_{\mathrm{eff}}/\kappa^2 \approx 0.07$ in the spectra of Fig.~\ref{fig:trans_spec_exp} represents a good compromise between this requirement and the condition of adiabaticity. In (a) spectra are plotted for $q = -2.8, N = 10^{5}$ and $p = -0.4, -1.3, -3.2$ (blue, green, orange trace). The values for $p$ are determined by measuring the power resonantly transmitted through the cavity with no atoms inside. This requires knowledge of the quantum efficiency of the detector, the transmission coefficient of the output mirror, and of subsequent optical losses. Hence, we believe that the absolute scale of measured $p$-values has an uncertainty of a factor two, while the relative values in different measurements are known to better than $5\, \%$. The value of $q$ is obtained by measuring the cavity resonance shift due to forward scattering. As seen in (b) these spectra are well reproduced by solutions of the full dynamical equations Eqs.~(\ref{eq:matter}) and (\ref{eq:light}) using the measured value $q = -2.8$, however for values  $p = -0.13, -0.43, -1.06$ about a factor 3 smaller than the values determined in the experiment, which could indicate that we overestimate the experimentally realized intra-cavity photon numbers. The factor three discrepancy disappears, if we assume a $q$-value in the calculations reduced by about $20\, \%$, however, at the price that the agreement of the resonance peak positions with those in the measurements is slightly degraded. The similarity of the curves in (a) and (b) to the $\Ww <0$ stationary solutions shown in Fig.~\ref{fig:steadystate}(a) is obvious. The situation is quite different for frequency scans starting on the blue side of the cavity resonance ($\dot \delta_{\mathrm{eff}} < 0$) with the atoms prepared in the upper state $\ket{\cos(2kx)}$. Atoms in $\ket{\cos(2kx)}$ can undergo collisional scattering, which populates a continuum of scattering states. In our experiment, due to the high density of several $10^{14}\,$cm$^{-3}$, more than half of the atoms are rapidly transferred to such states not accounted for by our two-mode description. The resulting spectra rather resemble those of thermal atomic samples than the $\Ww > 0$ stationary solutions shown in Fig.~\ref{fig:steadystate}(a). In Fig.~\ref{fig:trans_spec_exp}(c) we show $\dot \delta_{\mathrm{eff}} > 0$ scans for thermal samples prepared at about $210$~nK slightly above the critical temperature and close to the recoil temperature of about $170$~nK such that the atoms reside well within the first Brillouin zone $[-\hbar k, \hbar k]$. The atom number is $3 \times 10^{5}$ atoms with a correspondingly large cooperativity parameter of $q=-8.3$ and $p = -0.38, -0.49, -1.04$ (blue, green, orange trace). The steep rise in the left wings of the resonance features of these spectra clearly indicates their bistable character. This is emphasized in (d), which compares two spectra for an even larger ($N = 4.4 \times 10^{5}$, $q=-12.2$, $p=-1.47$) thermal sample and opposite scan directions, showing significant hysteresis. The frequency detunings, where the sudden rise of the intra-cavity intensity occurs, agree with the predictions of the $\Ww <0$ steady state solution of Eq.~(\ref{eq:stationary_states_b}) (black solid trace in (d)), although this solution relies on the two-mode description. This is in accordance with the expectation that deviations should only arise for samples at temperatures well above the recoil energy, such that higher bands are populated. 

\subsection{Cavity-induced momentum transfer}
\label{sec:momentum-transfer}
\begin{figure}[bth]
\centering
\includegraphics[scale=0.7, angle=0, origin=c]{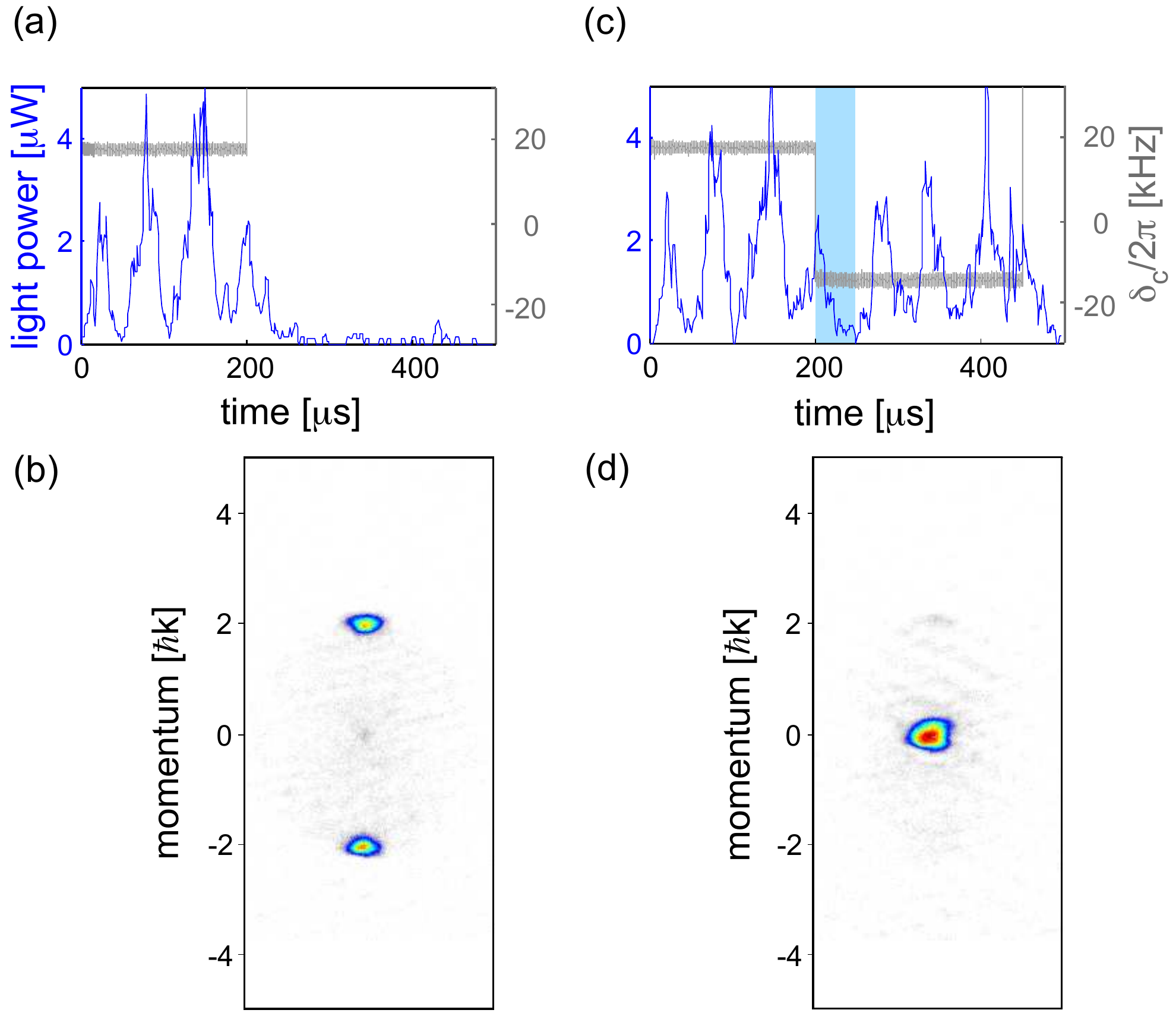}
\caption{Resonant momentum transfer. (a) Intra-cavity power for a $200\,\mu$s long heating pulse with positive detuning $\delta_{\mathrm{eff}}/\kappa = 4$ and $p = -5.6$. (b) Population of momentum classes produced by the pulse in (a). (c) and (d) correspond to (a) and (b), however, after an additional $200\,\mu$s long cooling pulse with negative detuning $\delta_{\mathrm{eff}}/\kappa = -3.1$ and $p = -5.6$. The light blue bar indicates a $40\,\mu$s long period with the pump beam blocked. The cooperativity parameter is $q=-2.2$.}
\label{fig:momentum_transfer_exp}
\end{figure}

In certain windows of the pump frequency superradiant back-scattering instabilities arise, which yield directional transfer of atoms between the states $\ket{0}$ and $\ket{\cos(2kx)}$. The elementary processes, sketched in Fig.~\ref{fig:momentum_transfer}(a) and (b), can be directly observed experimentally. Examples are shown in Fig.~\ref{fig:momentum_transfer_exp}. In (a) and (b), after preparation of the BEC in $\ket{0}$, a (heating) pump pulse of $200\,\mu$s duration is applied with positive effective detuning $\delta_{\mathrm{eff}}/\kappa = 4$. The power transmitted through the cavity (blue trace) recorded in (a), which monitors the intra-cavity power, shows oscillatory dynamics. This indicates the conversion of pump photons at frequency $\omega_{\mathrm{p}}$ into photons resonant with the cavity via back-scattering from the atomic sample, which yields a beat in the transmitted light at the difference frequency. After few cycles most of the atoms are transferred to the upper motional state $\ket{\cos(2kx)}$, which can be seen in (b), where the momentum distribution at the end of the pulse is recorded. This is accomplished by a band mapping technique, which maps quasi-momentum into momentum, i.e. the population of the n-th band is mapped onto the n-th Brillouin zone. In the experiment, the lattice is switched off with moderate speed determined by $\kappa$ and after a $\unit[25]{ms}$ ballistic flight the atomic density distribution is recorded. The plot in (b) shows that after the excitation pulse both momenta $\pm 2 \hbar k$ are equally populated, while the zero momentum class is entirely depleted. The grayish background arises because atoms in the different momentum states ($0, \pm 2 \hbar k$) elastically scatter into a continuum of scattering states. While these atoms (up to 60 percent) remain in the cavity and contribute to the atom-cavity coupling, their dynamics is not captured by the dual mode description applied throughout this work. Note, that we do not see atoms with momenta $\pm 2 n \hbar k$ with $n>1$, which shows that excitation into higher bands by multiple back-scattering is in fact suppressed due to the sub-recoil energy resolution of the cavity. In (c) and (d) about $40\,\mu$s after termination of the blue detuned excitation pulse discussed in (a) and (b) a red detuned (cooling) pulse of the same duration is applied ($\delta_{\mathrm{eff}}/\kappa = -3.1$), which transfers the atoms back to zero momentum. 

\subsection{Dynamics of the intra-cavity light field}
\label{sec:intra_cavity_dynamics}

\begin{figure}[bt]
\centering
\includegraphics[scale=0.7, angle=0, origin=c]{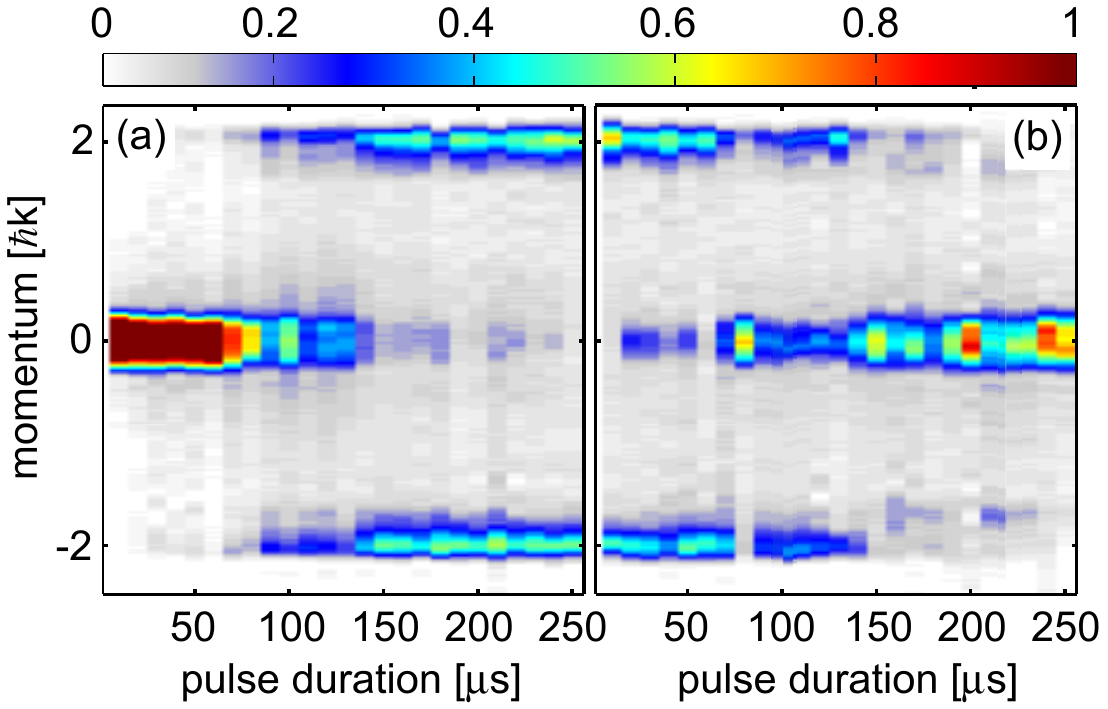}
\caption{Time evolution of the atomic momentum distribution for a heating pulse with $\delta_{\mathrm{eff}}/\kappa = 4$ and $p = -3.1$ in (a) and a subsequent cooling pulse with $\delta_{\mathrm{eff}}/\kappa = -3.1$ and $p = -2.3$ in (b). The cooperativity is $q=-2.2$ in both cases. The colors parametrize the observed optical density scaled to its maximal value.}
\label{fig:timeevolve}
\end{figure}

Observations as in Fig.~\ref{fig:momentum_transfer_exp} lead us to he following overall picture of the underlying dynamics. For sufficiently low pump strengths the resonance domains for heating and cooling pulses are well separated. The resonant photons emitted into the cavity mode leave the cavity at the rate $2 \kappa$ and hence the momentum transfer acquires a directionality. In fact, while on a time scale of a few photon lifetimes oscillatory dynamics is observed, after longer pulse durations the atomic sample undergoes complete transfer. If the pump strength increases, the resonance conditions are power broadened such that heating and cooling transfer eventually becomes possible for the same setting of the pump frequency. In this case persistent oscillatory behavior arises. In Fig.~\ref{fig:timeevolve} the time dependence for the case of low pump strength is studied. A series of vertical sections midway through plots as in Fig.~\ref{fig:momentum_transfer_exp}(b) is recorded versus the pump pulse duration for a heating pulse with $\delta_{\mathrm{eff}}/\kappa = 4$ and $p = -3.1$ in (a) and for a cooling pulse with $\delta_{\mathrm{eff}}/\kappa = -3.1$ and $p = -2.3$ in (b). In either case a complete transfer is seen, which requires minimal pulse durations of about $150\,\mu$s. The appearance of a minimal time required to start the momentum transfer reflects the fact that for the initial states ($\ket{0}$ or $\ket{\cos(2kx)}$) the back-scattering contributions of different atoms sum up destructively. Hence, back-scattering requires that initially thermal or quantum fluctuations provide a small transient spatial modulation of the density distribution commensurable with the resonance wavelength of the cavity. This initializes superradiant back-scattering and the momentum transfer progresses exponentially. This exponential growth saturates, if zero inversion is approached. The superradiant character of back-scattering vanishes again as $|\Ww|$ approaches unity again.

\begin{figure}[bt]
\centering
\includegraphics[scale=0.9, angle=0, origin=c]{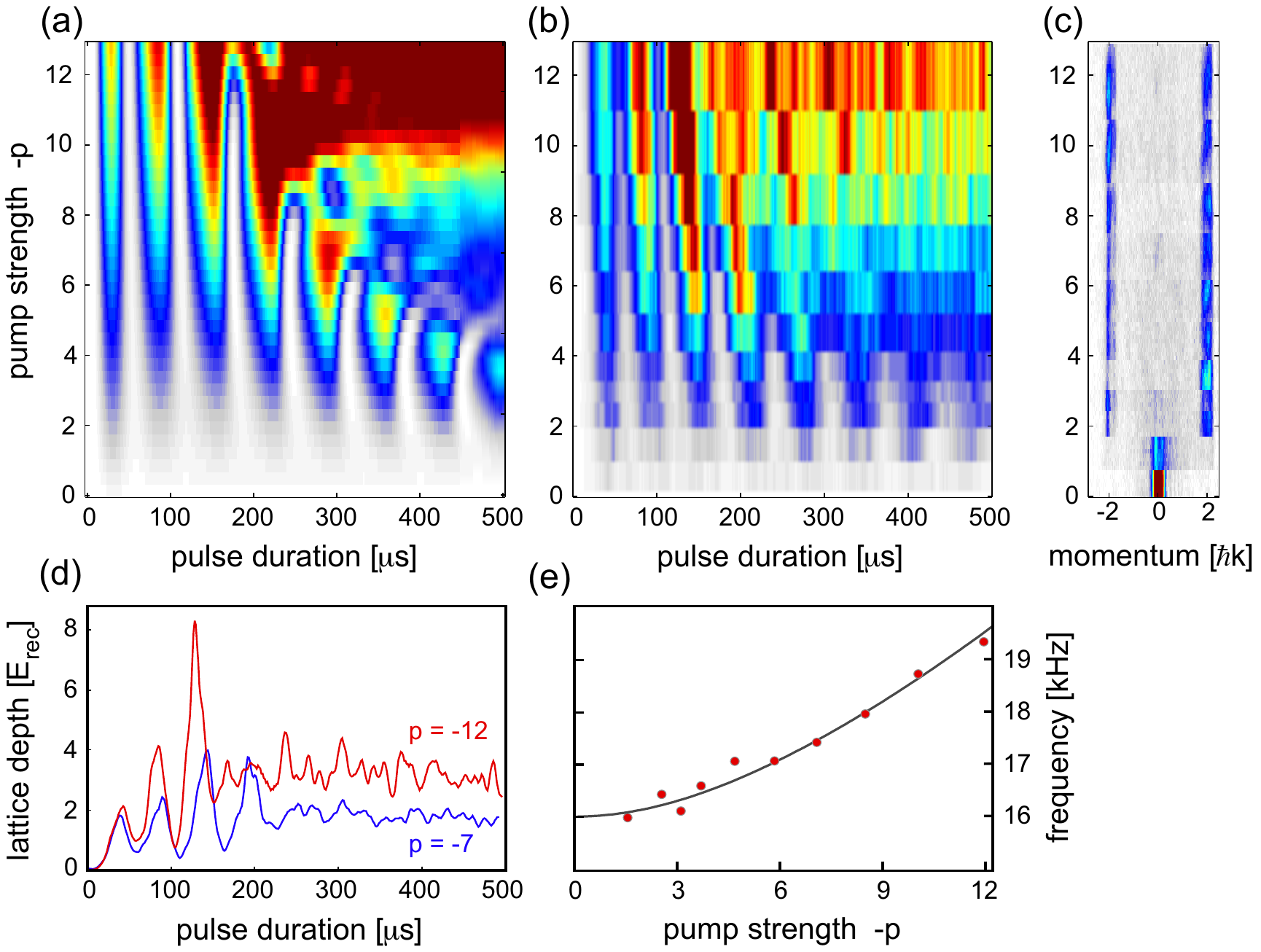}
\caption{The intra-cavity photon number is plotted versus pump strength and pump duration for a blue detuned pulse with $\delta_{\mathrm{eff}}/\kappa = 4.4$ and $q=-2.8$. Calculations using Eq.~(\ref{eq:matter}), (\ref{eq:light}) with $c_0 = 0$ (a) and observations (b) are shown. The colors in (a) and (b) indicate photon numbers between zero (white) and $3\times10^{4}$ (dark red). The calculations in (a) have been scaled down by a factor two, to attain optimal agreement with the observations in (b). (c) shows momentum spectra after a pulse of 0.5~ms duration corresponding to the right edge in (a) and (b). (d) Horizontal sections through (b) for -p = 12 (red) and -p = 7 (blue). (e) Frequencies of the initial oscillations in (d) plotted versus the intra-cavity lattice depth (red disks). The solid line shows the results of Eqs.~(\ref{eq:eigenvalues}) with $c_0 = 0.9$}
\label{fig:excitation_dynamics}
\end{figure}

In Fig.~\ref{fig:excitation_dynamics} the intra-cavity photon number is plotted versus the pump strength and the pump duration for a blue detuned pulse with $\delta_{\mathrm{eff}}/\kappa = 4.4$ and $q=-2.8$. Calculations based upon Eq.~(\ref{eq:matter}), (\ref{eq:light}) and corresponding observations are shown in (a) and (b), respectively. The main features visible in the data in (b) are qualitatively reproduced by the theory in (a) where collisional interaction is neglected. Note, however, that the calculated intra-cavity photon numbers in (a) have been scaled down by a factor two, to attain optimal agreement with the observations in (b). Similarly as discussed in the context of Fig.~\ref{fig:trans_spec_exp}, this indicates that our experimental determination overestimates the $q$-value on the $10 \,\%$ level. For each pulse duration a minimal pump strength is required to couple photons to the cavity and hence transfer atoms from $\ket{0}$ to $\ket{\cos(2kx)}$. This is confirmed in (c) where the momentum distribution of the atoms is plotted versus the pump strength for a pulse duration of 0.5~ms, which corresponds to the outermost right-hand columns in (a) and (b). After a minimal pump strength is reached, a nearly complete rapid transfer of the BEC into the momentum states $\pm 2 \hbar k$ is observed. The minimal pump strength approaches zero as the pump duration goes to infinity although very gradually. Hence, there is no true threshold even for large system sizes, as for example known to occur for transversally pumped cavities \cite{Dicke1, Dicke2, Dicke3, Dicke4} or in collective atomic recoil lasing scenarios \cite{CARL1, CARL2, CARL3}. The oscillatory dynamics visible in (a) and (b) provides further interesting insights. Horizontal sections through (b) are plotted in (d) for values of the pump strength $p = -12$ (red trace) and $p = -7$ (blue trace). At short times distinct transient oscillations in the kHz-domain are clearly seen showing that the intra-cavity light field comprises two frequency components, which give rise to a beat signal. One component is the injected pump light at frequency $\omega_{\mathrm{p}}$, which within the intra-cavity photon lifetime approaches a steady state level determined by the chosen value of $\delta_{\mathrm{eff}}$, thus building up a stationary intra-cavity light-shift potential. The second component arises from photons emitted into the cavity at a lower frequency $\omega_{\mathrm{p}} - \Omega$ in back-scattering processes exciting the atoms to $\ket{\cos(2kx)}$. After the transfer is completed this component and hence the oscillation terminates. 

The frequency difference $\Omega$ should reproduce the energy separation between $\ket{0}$ and $\ket{\cos(2kx)}$, which in absence of a lattice potential and for zero interaction should amount to $4 \hbar \omega_{\mathrm{rec}}$. If the lattice potential is accounted for, the states $\ket{0}$ to $\ket{\cos(2kx)}$ correspond to the zero quasi-momentum states in the first and third bands, with an energy separation exceeding $4 \hbar \omega_{\mathrm{rec}}$ by an amount depending on the lattice depth. An additional increase arises due to collisional interaction. This expectation is in fact confirmed by the observations in (e), where the oscillation frequencies - extracted from traces as shown in (d) - are plotted versus the pump strength. The solid line shows a calculation of the difference between the two eigenvalues of Eq.~(\ref{eq:Heisenberg}) accounting for an intra-cavity lattice and collisional interaction: Inserting $(\beta_0(t),\beta_2(t)) = (\beta_0^{(0)},\beta_2^{(0)}) \,e^{i \Omega t}$ into Eq.~(\ref{eq:Heisenberg}) yields the two equations
\begin{eqnarray}
\label{eq:eigenvalues0}
\left(\frac{1}{\sqrt{8}} \alpha^{\dagger} \alpha \, \Delta_0 + c_0 \Xx \right)\,\Ww = \frac{1}{2}   \left(4 \omega_{\mathrm{rec}} + \frac{1}{4} c_0 (1+\Ww) \right) \Xx
\nonumber \\
\Omega = \left(\frac{1}{2} \alpha^{\dagger} \alpha \, \Delta_0 + c_0 \right) + \frac{1}{2}   \left(4 \omega_{\mathrm{rec}} + \frac{1}{4} c_0 (1+\Ww) \right) \frac{(1+\Ww)}{\Ww}\, .
\label{eq:eigenvalues}
\end{eqnarray}
The upper line in Eq.~(\ref{eq:eigenvalues}) together with the normalization condition $1= \Ww^2+\Xx^2$ allows one to evaluate $\Ww$ as a function of the intra-cavity lattice well depth $\alpha^{\dagger} \alpha \, \Delta_0$. This may be inserted into the lower line to obtain $\Omega(\alpha^{\dagger} \alpha \, \Delta_0)$. The solid line shown in (e) is obtained for $c_0 = 0.9$ and a maximal lattice depth of $4 \hbar \omega_{\mathrm{rec}}$ corresponding to $p = -12$. The calculation of $\Omega$ appears to agree well with the observations, however, the line density corresponding to $c_0 = 0.9$ is nearly three times larger than what a three-dimensional Thomas-Fermi density distribution predicts for $10^5$ particles and the geometry of our external trap. We attribute this to the simplifying assumption at the basis of Eq.~(\ref{eq:2modemodel}) that the transverse density distribution is fixed, independent of $\rho_{\mathrm{1D}}$, which becomes questionable, if $a_s\,\rho_{\mathrm{1D}} >1$ as in our trap \cite{Ols:98, Sal:02, Mun:08}.

\subsection{Back-scattering by a moving BEC}
\label{sec:moving_BEC}
\begin{figure}[bt]
\centering
\includegraphics[scale=0.6, angle=0, origin=c]{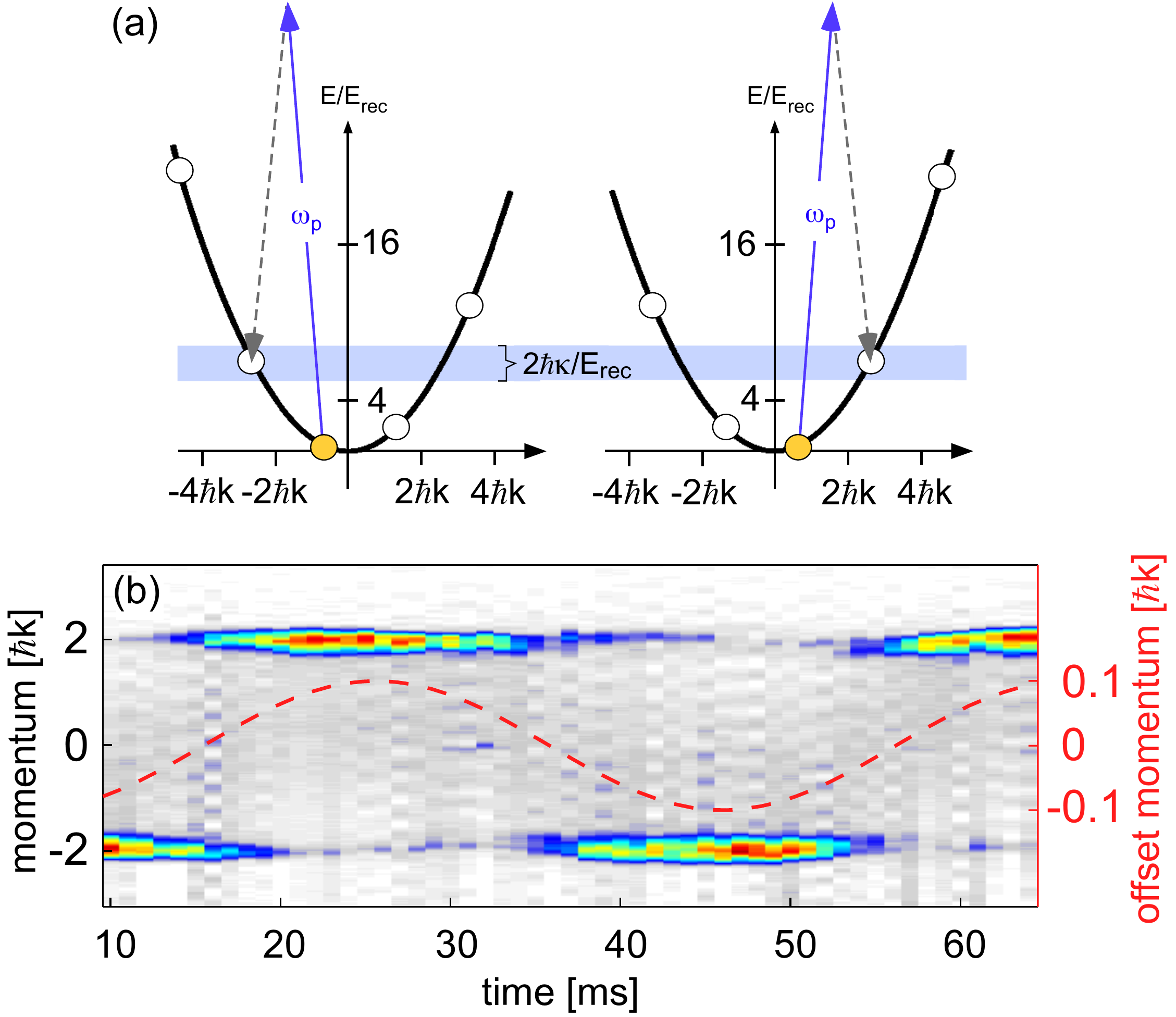}
\caption{Population oscillation as a result of a small offset momentum. (a) Single particle dispersion relation with accessible momentum states including a negative (left panel) or positive (right panel) offset momentum of the BEC. (b) Momentum distributions observed subsequent to a $0.4$~ms long blue detuned excitation pulse ($\delta_{\mathrm{eff}}/\kappa = 4.2$, $q=-2.8$, $p=-1$) for varying initial momenta of the BEC quantified by the red dashed trace.}
\label{fig:oscillation}
\end{figure}

In this section we consider atomic samples with non-vanishing center-of-mass (CM) momentum with respect to the cavity mode. Because of the narrow transmission linewidth of the cavity, the population of the $\pm 2 \hbar k$ momentum states after application of a heating pulse sensitively depends on the initial CM momentum. This is illustrated in Fig.~\ref{fig:oscillation}(a) where the parabolic single particle dispersion is plotted together with the momentum states accessible via resonant back-scattering, if an initial negative (left panel) or positive (right panel) CM momentum is assumed. To verify this experimentally, the BEC is slightly kicked, such that it performs an oscillation along the cavity axis in the magnetic trap with a frequency of $\omega_z/2\pi\approx\unit[25.2]{Hz}$. Hence, different CM momenta are realized for different trap holding times. In Fig.~\ref{fig:oscillation}(b) a momentum swing is chosen to cover the interval $\pm 0.1\, \hbar k$, indicated by the red dashed trace, and after a variable trap holding time a heating pulse is applied and the resulting momentum distribution is recorded. Since the pulse duration ($\tau_{\mathrm{pulse}}=\unit[200]{\mu s}$) is chosen much smaller than the oscillation period ($\tau_{\mathrm{osc}}\approx \unit[40]{ms}$) the change of the initial momentum during the excitation may be neglected. As is seen in (b), a complete oscillation of the population between the states $2 \hbar k$ and $- 2 \hbar k$ is observed. We are thus able to detect initial CM momenta to better than a percent of the recoil momentum $\hbar k$. For a CM momentum of $p_{\mathrm{cm}} = \pm 0.1 \,\hbar k$ the resonance conditions for a transfer to the $\pm 2 \hbar k \pm p_{\mathrm{cm}}$ states differ by only $0.8 \,\hbar k$, which is significantly smaller than the cavity linewidth. Nevertheless, the system appears to favour population of a single momentum state, which we interpret as a signature of bosonic enhancement.

\section{Conclusion and Outlook}
\label{sec:conclusion-outlook}
In this work we have presented an extensive study of a Bose-Einstein condensate (BEC) interacting with a single longitudinal mode of a standing wave resonator, which is coupled along the cavity axis by a weak external laser beam. This most elementary scenario of atom-cavity physics has been considered in many previous investigations, however, not in the most fundamental regime combining cavity dominated scattering, strong cooperative coupling, and sub-recoil resolution. Focussing here on this novel regime, we have unveiled rich non-linear mean field dynamics with typical signatures as bistability, hysteresis, persistent oscillations, and superradiant back-scattering instabilities. Future studies focussing on the role of quantum fluctuations may allow to discover significant quantum entanglement between light and matter observables.

\ack
This work was partially supported by DFG-He2334/14-1, DFG-SFB 925, DFG-GrK1355. We are grateful for useful discussions with Michael Thorwart and Reza Bakhtiari.

\end{document}